\documentclass[prb,twocolumn,showpacs,preprintnumbers,superscriptaddress]{revtex4-1}
\usepackage[usenames,dvipsnames]{color}
\usepackage{amssymb} 
\usepackage{times}
\usepackage{graphicx}
\usepackage{graphicx}
\usepackage{dcolumn}
\usepackage{datetime}
\usepackage{soul}
\usepackage{subfigure}
\usepackage[bookmarks, colorlinks=true, plainpages = false,citecolor = red, urlcolor = black, 
filecolor = blue]{hyperref}
\usepackage{multirow} 
\usepackage[english]{babel}
\usepackage{graphicx}
\usepackage{amsmath}
 
\usepackage[english]{babel}
\usepackage{graphicx}
\usepackage{amsmath}
 
\begin{document}
\title{Rashba-Dirac cones at the tungsten surface: Insights from a tight-binding model 
and thin film subband structure}
 
\author{George Kirczenow} 

\affiliation{Department of Physics, Simon Fraser
University, Burnaby, British Columbia, Canada V5A 1S6}

\date{\today}

\begin{abstract}\noindent
Phys. Rev. B94, 205414 (2016). Published Nov. 10, 2016.\\
URL: http://link.aps.org/doi/10.1103/PhysRevB.94.205414\\
DOI: 10.1103/PhysRevB.94.205414\\ 

\noindent A tight-binding model of bcc tungsten that includes spin-orbit
coupling is developed and applied to the surface states
of (110) tungsten thin films. The model describes accurately 
the anisotropic Dirac cone-like dispersion
and Rashba-like spin polarization of the surface states, including the crucial effect of the 
relaxation of the surface atomic layer of the tungsten towards the bulk. It is shown that the
surface relaxation affects the tungsten surface states because it results in increased overlaps 
between atomic orbitals of the surface atomic layer and nearby layers whereas electric fields that
are due to charge transfer between the tungsten and the vacuum near
the surface or between the bulk and surface layers do not significantly affect the
Rashba-Dirac surface states. It is found
 that hybridization with bulk modes has differing strengths 
 for thin film surface states belonging to the upper and lower Rashba-Dirac 
 cones and results in {\em reversal} of the directions of travel 
 of spin $\uparrow$ and $\downarrow$ 
electrons in most of the upper Rashba-Dirac 
 cone relative to those expected from phenomenology. 
 It is also shown that intrasite (not intersite) matrix elements
of the spin-orbit Hamiltonian are primarily responsible for the formation of the 
Rashba-Dirac cones, and their spin polarization. This finding should be considered
when modeling topological insulators, the spin Hall effect and related phenomena.

\end{abstract}

\maketitle
\section{Introduction}

Spin-orbit coupling in non-magnetic materials gives rise to spin-split  electronic
states in Rashba systems\cite{BR1,BR2} and topological insulators.\cite{KM1,KM2} 
The resulting locking of spin to the 
crystal-momentum offers potential avenues for the realization 
of novel spintronic devices and quantum computation. 
Electronic states that are spin-split due to strong spin-orbit coupling also occur 
at the surfaces of some non-magnetic metals, including 
gold,\cite{Au1,Au2,Au3,Au4,AuAg}
tungsten,\cite{Rotenberg1998,Hochstrasser2002,Shikin2008,Miyamoto2012a,
Miyamoto2012b,Rybkin2012,Biswas2013,Mirhosseini2013a,Giebels2013,
Mirhosseini2013b,
Braun2014,Miyamoto2015,Wortelen2015,Miyamoto2016} 
silver,\cite{Au2,AuAg} copper,\cite{Cu1,Cu2} bismuth,\cite{Bi1,Bi2,Bi3,Bi4,Bi5,Bi6} 
antimony,\cite{Sb} and iridium.\cite{Ir} 

Following the recent experimental observation at the tungsten (110) 
surface of 
Rashba-like spin-split and spin-polarized 
electronic states forming an anisotropic Dirac cone-like band 
structure,\cite{Miyamoto2012a}
there has been renewed interest in developing a better understanding of this 
system.\cite{Miyamoto2012a,
Miyamoto2012b,Rybkin2012,Biswas2013,Mirhosseini2013a,Giebels2013,
Mirhosseini2013b,
Braun2014,Miyamoto2015,Wortelen2015,Miyamoto2016} It has been 
demonstrated\cite{Miyamoto2012b}
that the experimentally observed anisotropic Dirac cone-like band 
structure can be fitted accurately by a phenomenological third order Rashba
model for surfaces with $C_{2v}$ symmetry.\cite{Vajna} {\em Ab initio}
calculations\cite{Rybkin2012,Mirhosseini2013a,Mirhosseini2013b,Braun2014,
Wortelen2015,Miyamoto2016} have also accounted for the 
experimental 
data, including the Dirac cone-like dispersion and the
spin polarizations of the observed surface states. 
However, it is also of interest to investigate 
the underlying physics with the help of a tight-binding model that, unlike 
the anisotropic Rashba phenomenology,\cite{Miyamoto2012b} is
atomistic and also can provide insights that are complimentary to those  
obtained from  
{\em ab initio} 
calculations.\cite{Rybkin2012,Mirhosseini2013a,Mirhosseini2013b,Braun2014,
Wortelen2015,Miyamoto2016} Such a tight-binding
model is introduced here and applied to tungsten thin films with
(110)  surfaces and their electronic subband structure.

The present tight-binding model is a modification of extended H\"{u}ckel 
theory,\cite{Wolfsberg,Hoffman,Ammeter,review2010}
generalized to include spin-orbit coupling as was proposed in 
Refs. \onlinecite{PRBrapid,PRB} for molecular nanomagnets. It is
parameterized to accurately reproduce the electronic band structure 
of {\em bulk} b.c.c. tungsten, both with\cite{Christensen,Glantschnig}
and without\cite{Papaconstantopoulos} 
the inclusion 
of spin-orbit coupling. The results obtained for the W(110) Rashba-Dirac
cone surface states are in remarkably good agreement with
those of the previous {\em ab initio}\cite{Rybkin2012,
Mirhosseini2013a,Mirhosseini2013b,
Braun2014,Miyamoto2016} calculations for semi-infinite
tungsten crystals. However, the 
present tight-binding approach and investigation of the thin film
subband structure yields additional insights outlined below. 

Although the present model is parameterized by fitting to {\em only 
the bulk} electronic structure of tungsten, it also captures correctly the  
important effect\cite{Mirhosseini2013a} on the Rashba-Dirac cone 
surface states of the relaxation of the 
W(110) surface atomic layer towards the bulk
that is observed experimentally.\cite{Meyerheim,Venus}
It is shown below that this effect is 
due to the increased overlaps between atomic
orbitals of the tungsten surface and bulk atomic layers induced by the
relaxation. It is also shown that the contribution of electric fields that
are due to charge transfer between the tungsten and the vacuum near
the surface or between the bulk and surface layers to this effect and 
their influence on the Rashba-Dirac cone-like 
dispersion of tungsten (110) surface 
states are both at most minor.

The present work also reveals that hybridization with bulk modes results
in the directions of travel of spin $\uparrow$ and $\downarrow$ electrons 
in the thin film's surface states belonging to most of the 
upper Rashba-Dirac cone 
being reversed relative to those predicted by    
phenomenological Rashba models.
This reversal is due to dense anticrossings
between the surface and bulk-like states that arise from their strong hybridization.
It is also shown that hybridization between the bulk-like states
and lower Rashba-Dirac cone states is much weaker than that
between the bulk states
and upper Rashba-Dirac cone states, suggesting that the intrinsic
lifetimes of the upper and lower Rashba-Dirac cone surface states 
may differ significantly. 

Finally, a comparison of the present model with the standard tight
binding formulation\cite{Mireles,KM2} of the original Rashba model\cite{BR1}
is made and strongly suggests that
the mechanism responsible for the Rashba-like dispersion and spin
polarization of W(110) surface states differs fundamentally from that
of the original Rashba effect in nearly free 2DEG's at semiconductor
interfaces. This conclusion is consistent 
with that reached previously by entirely different 
reasoning\cite{Krasovskii} for surface states of other 
materials.
It may be expected to have significant implications 
for theoretical work on topological
insulators, spin Hall phenomena and related topics.

The remainder of this article is organized as follows: 
The tight-binding model of tungsten is explained
in Section \ref{M}. The results obtained by applying 
the model to tungsten (110)
thin films are presented and discussed in Section \ref{Results}. 
The conclusions
drawn from this work are summarized in Section \ref{Discussion}.
  
\section{Model} 
\label{M}
The tight binding model developed here is based on 
extended H{\"u}ckel theory,\cite{Wolfsberg,Hoffman,Ammeter,review2010} 
a semi-empirical tight
binding scheme from quantum chemistry that is 
formulated in terms of a 
small set of  Slater-type atomic valence orbitals $\{
|\phi_i\rangle \}$, their overlaps $O_{ij} =
\langle\phi_i | \phi_j\rangle$ and a Hamiltonian matrix
$H^0_{ij} =
\langle\phi_i |H^0| \phi_j\rangle$. In extended H{\"u}ckel theory the diagonal
Hamiltonian elements $H^0_{ii} = \epsilon_i$  are chosen to
be the (negative) atomic orbital ionization energies. 
The Hamiltonian matrix elements 
for $i \ne j$ are approximated by
\begin{equation}
H^0_{ij} = K_{ij}
O_{ij}(\epsilon_i + \epsilon_j)/2
\label{Kij}
\end{equation}
In the Wolfsberg-Helmholz form of extended H{\"u}ckel 
theory,\cite{Wolfsberg} the empirical parameters $K_{ij}$ are
 all set to 1.75 in order for the model to yield approximate 
 energy levels for a variety of simple molecules. In the
 present work the $\epsilon_i$ and $K_{ij}$ are fitting 
 parameters chosen so that 
the model accurately reproduces the non-relativistic band structure of bulk 
bcc tungsten given in Ref. \onlinecite{Papaconstantopoulos} 
if the Hamiltonian matrix is given by equation (\ref{Kij}) and
the overlap matrix $O_{ij}$ is calculated using a standard
extended H{\"u}ckel software package.\cite{YAEHMOP} 
First, second and third neighbor matrix elements $H^0_{ij}$ and $O_{ij}$
are included in the present model. The values of the parameters
$\epsilon_i$ and $K_{ij}$ used in this work are given in 
Tables \ref{tab:1} and \ref{tab:2}. With these parameter values,
the present model (without spin-orbit coupling) 
matches the non-relativistic band structure of bulk 
bcc tungsten given in Ref. \onlinecite{Papaconstantopoulos}
well throughout the Brillouin zone in the energy range from $\sim$10 eV
below the Fermi level to $\sim$10 eV
above the Fermi level.

It is worth noting that in the tight-binding
model presented in Ref. \onlinecite{Papaconstantopoulos}, {\em all} of the
matrix elements $H^0_{ij}$ and $O_{ij}$ are treated as fitting parameters.
An advantage of the present methodology is that here, unlike in  
Ref. \onlinecite{Papaconstantopoulos}, the overlap 
matrix elements $O_{ij}$ are not fixed but depend on the 
atomic geometry of the tungsten
and hence the present formalism can treat deformations of the tungsten
crystal whereas the tight-binding model in Ref. \onlinecite{Papaconstantopoulos}
cannot. This is important since the relaxation of the W(110) surface atomic layer
towards the bulk\cite{Meyerheim,Venus} strongly affects the 
surface states\cite{Mirhosseini2013a} 
that are the topic
of the present work.

\begin{table}[t]
\caption{Tight-binding orbital energy parameters for bcc tungsten used in the
present work.
The $x,y$ and $z$ axes are aligned with the cubic crystal axes of bcc tungsten.}
\begin{center}
\begin{tabular}{c|c|c|c|c}
 Orbital & 6s & 6p & 5d$_{x^2-y^2}$, 
 5d$_{z^2}$ & 5d$_{xy}$,5d$_{xz}$,5d$_{yz}$\\ 
 \hline
 $\epsilon_i$ (eV) &  -11.907697 & -4.855731 &  -11.271223 & -10.365356\\
\end{tabular}
\end{center}
\label{tab:1}
\end{table}

\begin{table}[t]
\caption{Tight-binding parameters $K_{ij}=K_{ji}$ for bcc tungsten.
$\alpha,\alpha' = x, y ~\text{or}~z$. $\beta = xy, xz ~\text{or}~yz$.
$\gamma = x^2-y^2 ~\text{or}~z^2$. $\delta, \delta' = xy, xz, yz, x^2-y^2 ~\text{or}~z^2$.
The $x,y$ and $z$ axes are aligned with the cubic crystal axes of bcc tungsten.
The tungsten lattice parameter is $a=3.16$\AA.
}
\begin{center}
\begin{tabular}{l|c|c|c}
 neighbor & first & second & third\\ 
 \hline
 $K_{s,s}$ &  2.15 & 1.50 &  2.25\\
  \hline
 $K_{p_\alpha,p_\alpha}$ &  1.75 & 3.00 &  3.00\\
  \hline
 $K_{p_\alpha , p_{\alpha'}}$, $_{\alpha \ne \alpha'}$ &  2.30 & 2.00 &  2.00\\
   \hline
 $K_{d_\beta,d_\beta}$ &  2.30 & 1.70 &  1.70\\
   \hline
 $K_{d_\gamma,d_\gamma}$ &  2.00 & 2.00 &  2.00\\
  \hline
 $K_{d_\delta , d_{\delta'}}$, $_{\delta \ne \delta'}$ &  1.95 & 2.00 &  2.00\\
  \hline
 $K_{s,p_\alpha}$ &  2.75 & 2.15 &  2.25\\
   \hline
 $K_{s,d_\delta}$ &  2.25 & 2.30 &  2.40\\
  \hline
 $K_{p_\alpha , d_{\gamma}}$ &  2.40 & 2.00 &  3.50\\
  \hline
 $K_{p_\alpha , d_{\beta}}$ &  1.75 & 2.00 &  2.00\\

\end{tabular}
\end{center}
\label{tab:2}
\end{table}

$H^0_{ij}$, like standard extended H{\"u}ckel theory, 
does not include spin-orbit coupling which
is included in the present tight binding model 
using the formalism developed in  
Refs. \onlinecite{PRBrapid,PRB} to treat spin-orbit coupling
in molecular nanomagnets. As shown in Refs. \onlinecite{PRBrapid,PRB},
the spin-orbit Hamiltonian 
$H_{\mbox{\scriptsize{SO}}}=\frac{\hbar}{(2mc)^2}\boldsymbol{\sigma}\cdot  
\nabla{V(\bf{r}) \times \mathbf{p}  } $
can be approximated as a sum of atomic contributions and its 
tight-binding matrix elements can then be expressed 
as intrasite and intersite terms. The intrasite matrix elements for 
site $\alpha$ are\cite{PRBrapid,PRB}
\begin{equation}
 \langle \Psi_{\alpha l d s}|H^\text{intra}_{\mbox{\scriptsize{SO}}}| 
 \Psi_{\alpha l' d' {s}'} \rangle = \frac{\zeta_{l \alpha}}{\hbar^2} \langle 
 \alpha l d s | \mathbf S \cdot \mathbf L_{\alpha} | \alpha l d' {s}' \rangle\delta_{ll'}  
\label{intra}
\end{equation}
where
\begin{equation}
\zeta_{l \alpha}=\langle R_{\alpha l}  | \frac{1}{2m^2c^2} 
\frac{1}{|\mathbf{r}-\mathbf{r}_{\alpha}|} 
\frac{dV_\alpha(|\mathbf{r}-\mathbf{r}_{\alpha}|)}{d(|\mathbf{r}-\mathbf{r}_{\alpha}|)}  
| R_{\alpha l}  \rangle
\label{intra'}
\end{equation}
and the atomic orbital $|\Psi_{\alpha l d s}\rangle$ for site 
$\alpha$ is the product of a radial part $|R_{\alpha l}\rangle$ and 
directed atomic orbital $|\alpha l d s\rangle$. Here $l$ is the orbital angular 
momentum quantum number, $d$ may be $s, p_x, p_y, p_z, d_{xy},d_{xz},...$ 
depending on the value of $l$, and $s$ is the spin quantum number. The principal 
quantum number $n$ is suppressed. $\mathbf S$ is the spin angular momentum, 
$\mathbf L_{\alpha}$ is the orbital angular momentum relative to the 
nucleus located at $\mathbf{r}_{\alpha}$ and $V_\alpha$ is the electron 
potential energy contribution of atom $\alpha$. The
intersite matrix elements are \cite{PRBrapid,PRB}
\begin{equation}
\begin{split}
 \langle \Psi_{\alpha' l' d' s'}|H^\text{inter}_{\mbox{\scriptsize{SO}}}| 
 \Psi_{\alpha l d {s}} \rangle = (1-\delta_{\alpha' \alpha})\times\\  
 \sum_{d'' {s}''} [O_{\alpha' l' d' s',\alpha l d'' {s}''}\langle \Psi_{\alpha 
 l d'' {s}''}|H^\text{intra}_{\mbox{\scriptsize{SO}}}| \Psi_{\alpha l d {s}} \rangle\\
+ O_{\alpha l d s,\alpha' l' d'' {s}''}\langle \Psi_{\alpha' l' d'' s''}|
H^\text{intra}_{\mbox{\scriptsize{SO}}}| \Psi_{\alpha' l' d' {s}'} \rangle^*] 
\label{inter}
\end{split}
\end{equation}
where $ O_{\alpha l d s,\alpha' l' d' {s}'} = \langle \Psi_{\alpha l d s}| 
\Psi_{\alpha' l' d' {s}'} \rangle$ is the overlap between the valence
orbitals on sites $\alpha$ and $\alpha'$. 

The matrix elements of $\mathbf S \cdot \mathbf L/\hbar^2$ that appear in 
Eq. (\ref{intra}) have been evaluated analytically. 
Explicit expressions for them in the cubic harmonics representation
for the angular parts of $s, p$ and $d$ orbitals that is used in the YAEHMOP
extended 
H{\"u}ckel software package\cite{YAEHMOP} are given in 
Table II of Ref. \onlinecite{Konschuh}. They
do not depend on which material is being considered since only the
angular parts of the atomic orbitals are involved; the effects of the
radial parts are included in the parameters $\zeta_{l \alpha}$ defined by 
Eq. (\ref{intra'}). The parameters
$\zeta_{l \alpha}$ in Eq.  (\ref{intra}) and (\ref{intra'}) characterize the 
strength of the spin-orbit coupling. In the
present work their values are determined by fitting the bulk tungsten 
band structure obtained for the complete tight-binding
Hamiltonian $H=H^0 + H^\text{intra}_{\mbox{\scriptsize{SO}}}
+H^\text{inter}_{\mbox{\scriptsize{SO}}}$ (with $H^0$
parameterized according to Tables \ref{tab:1} and \ref{tab:2}) 
to the relativistic {\em ab initio} band structures for
bulk tungsten in Refs. \onlinecite{Christensen,Glantschnig}. This yields 
$\zeta_{5d \alpha} = 0.311$eV and $\zeta_{6p \alpha} =1.74$eV. 
These
values were obtained by fitting to the {\em ab initio} 
values\cite{Christensen,
Glantschnig} of band splittings
opened by the spin-orbit coupling at the $\Gamma$ point 
(specifically, the gap centered
$\sim 1$eV below the Fermi level) and H point (the gap centered
$\sim 8.5$eV above the Fermi level)
of the Brillouin zone. The size of the former gap is affected mainly by 
$\zeta_{5d \alpha}$
and the latter mainly by $\zeta_{6p \alpha}$.  With these parameter values,
the band structure obtained from the present model (with spin-orbit coupling) 
matches the relativistic band structures of bulk 
bcc tungsten given in Refs. \onlinecite{Christensen,Glantschnig}
reasonably well throughout the Brillouin zone in the energy range 
from $\sim$10 eV
below the Fermi level to $\sim$10 eV
above the Fermi level. 
\section{Results} 
\label{Results}

 \begin{figure*}[t]
\centering
\includegraphics[width=1.0\linewidth]{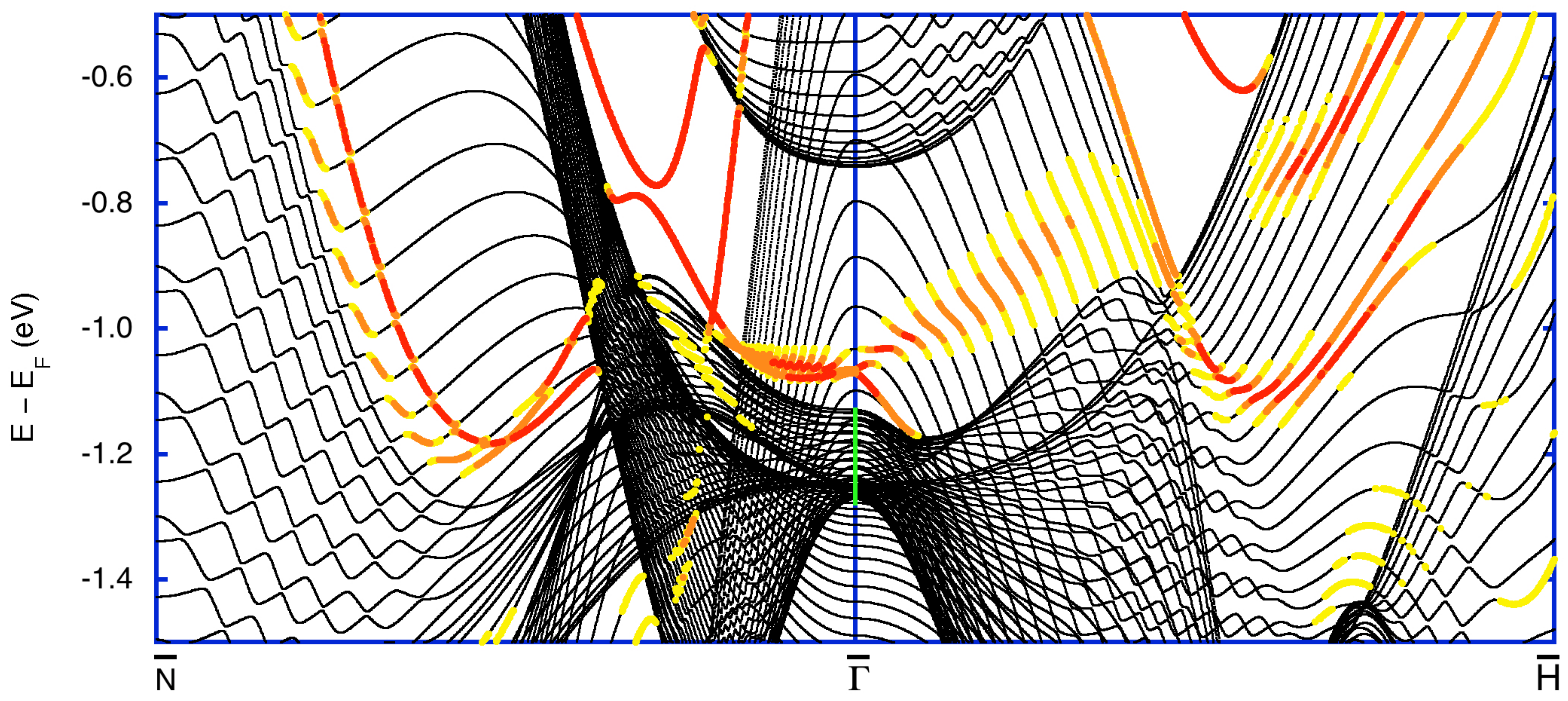}
\caption{
(Color online) Calculated energies $E$ of subband 
Bloch states vs. in-plane wave vector for a 50 atomic layer 
thin film of tungsten with (110) surfaces for high symmetry 
directions in the 2D Brillouin zone.
The surface atomic layers of the film are relaxed towards 
the bulk consistent with experiment.
\cite{Meyerheim,Venus}
States with larger probabilities $P$ of the electron being 
in the (110) surface atomic
layer are shown in color, $P > 0.3$ in red, $0.3 > P > 0.15$ 
in orange, $0.15 > P > 0.075$ in yellow.
Results for the opposite $(\bar{1}\bar{1}0)$ surface layer are similar. 
The upper and lower anisotropic Rashba-Dirac cones come 
together at their common apex at
the $\bar{\Gamma}$ point. The green line marks the energy 
range with a high density of
bulk-like states at the $\bar{\Gamma}$ point.
}
\label{surf} 
\end{figure*}

  \begin{figure}[t!]
\centering
\includegraphics[width=1.0\linewidth]{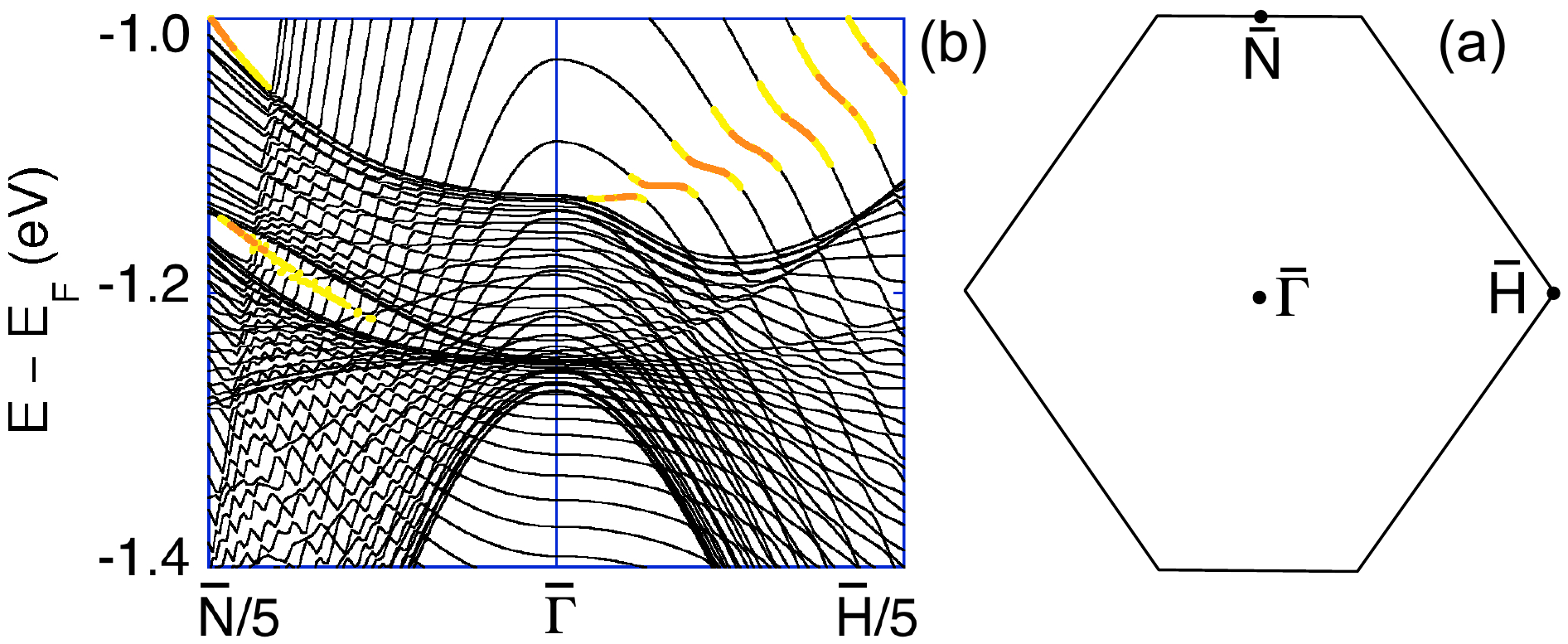}
\caption{
(Color online) (a) Brillouin zone of (110) tungsten film. (b) Dispersion 
near apex of Rashba-Dirac cone for 50 atomic layer 
thin film of tungsten with (110) surfaces and no surface relaxation.
Colored lines indicate states with larger probabilities of the electron 
being in the (110) surface atomic
layer. Notation as in Fig. \ref{surf}. Of the Rashba-Dirac cones
only a part of the upper cone in the $\bar{\Gamma}{-}\bar{\text{H}}$ direction
(shown in orange and yellow at the upper right) is visible. 
}
\label{BZN} 
\end{figure}

 \begin{figure*}[t]
\centering
\includegraphics[width=1.0\linewidth]{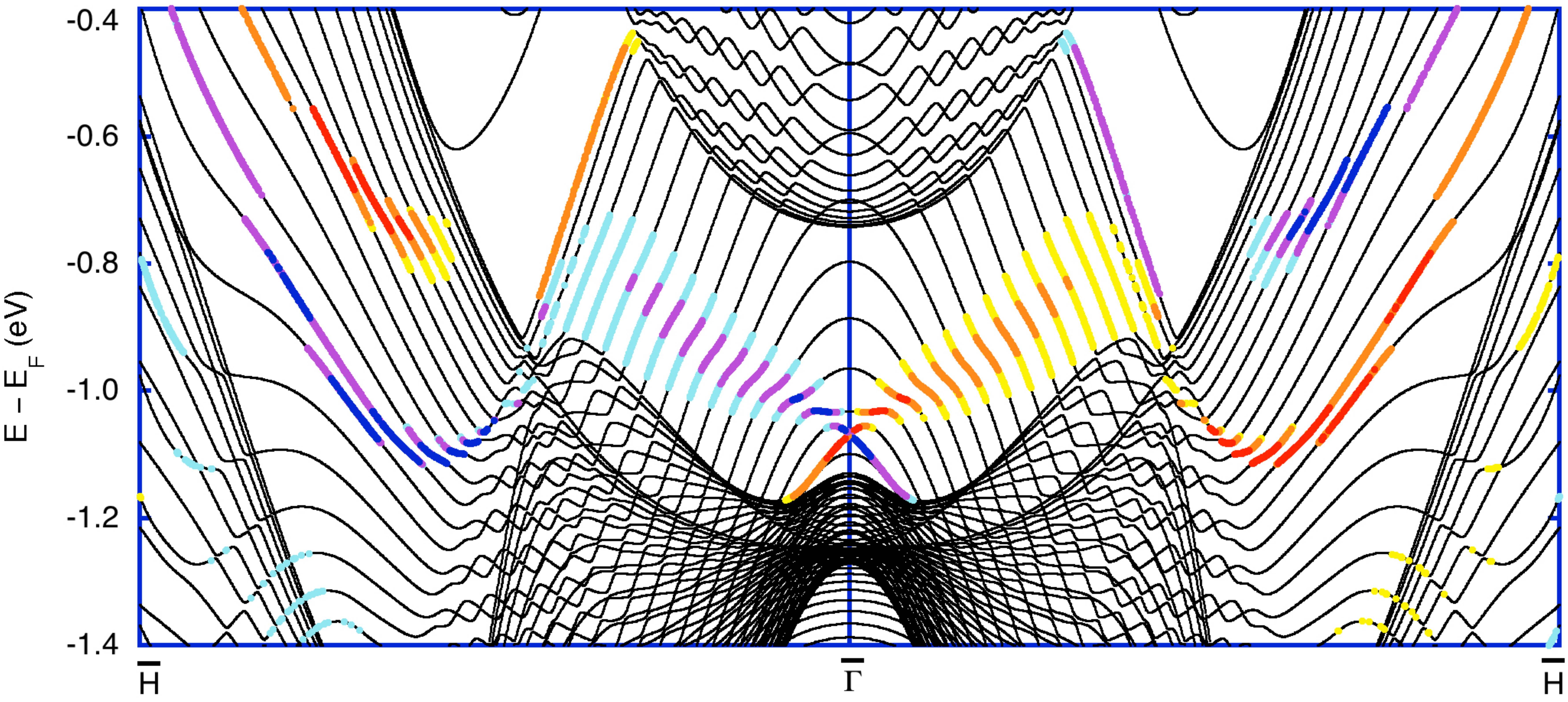}
\caption{
(Color online) Surface states with strong spin polarization (shown 
in color) in the (110) surface atomic layer of a 50 atomic layer 
thin film of tungsten with relaxed (110) surfaces. Calculated 
energies $E$ of subband Bloch states
vs. wave vector are plotted for a cut through the 2D Brillouin 
zone in the $\bar{\text{H}}$
direction. The spin quantization axis is in the (110) plane and 
perpendicular to the wave vector
of the Bloch state. 
States with more than 90\% spin $\uparrow$ 
in the (110) surface atomic layer and with probabilities $P$ 
of the electron being in the (110) surface atomic
layer are shown in red, orange and yellow for 
$P > 0.3$, $0.3 > P > 0.15$ and $0.15 > P > 0.075$ respectively. 
States with more than 90\% spin $\downarrow$
in the (110) surface atomic layer and with probabilities $P$ 
of the electron being in the (110) surface atomic
layer are shown in dark blue, mauve and pale blue for 
$P > 0.3$, $0.3 > P > 0.15$ and $0.15 > P > 0.075$ respectively.
}
\label{spin} 
\end{figure*}

 \begin{figure*}[t]
\centering
\includegraphics[width=1.0\linewidth]{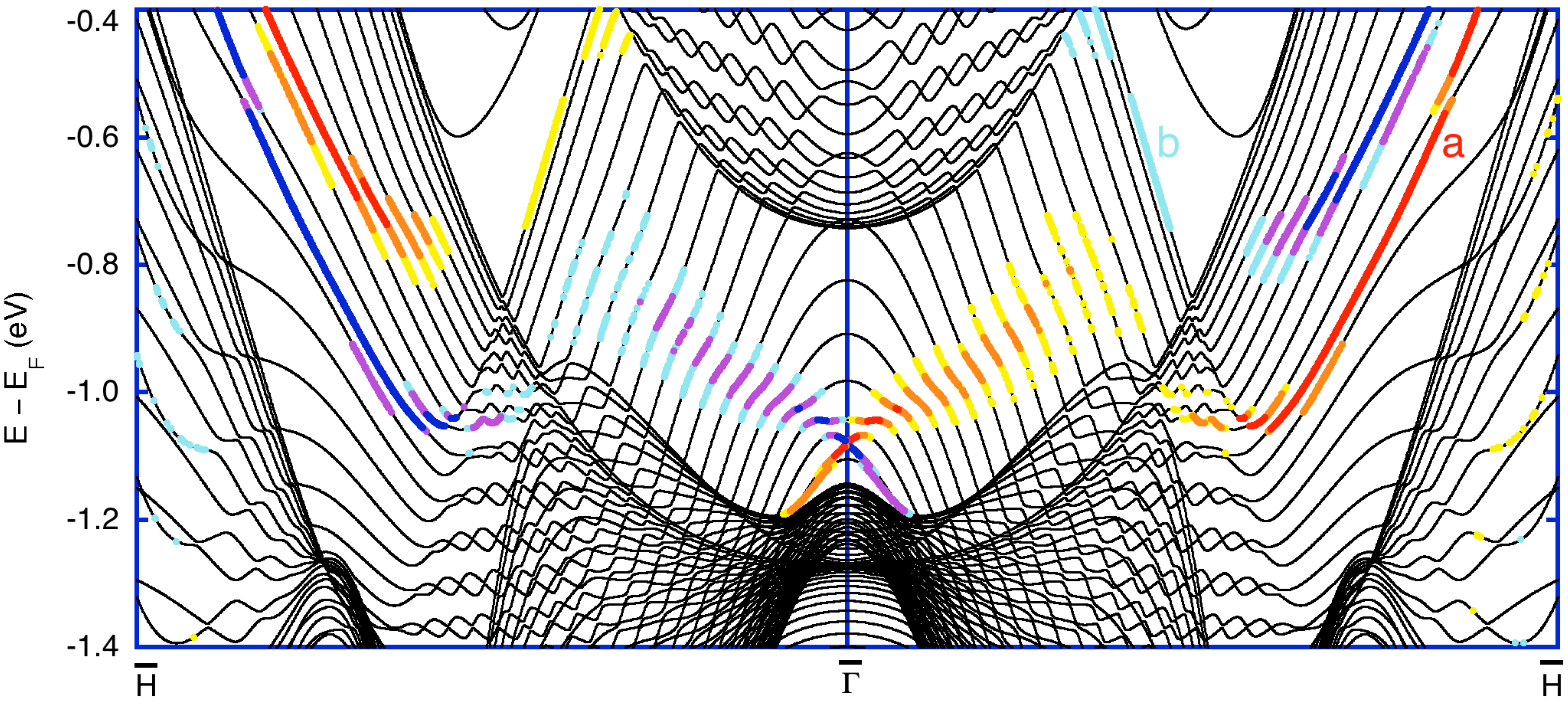}
\caption{
(Color online) Surface states with strong spin polarization 
(shown in color) in the (110) surface atomic layer of a 50 atomic layer 
thin film of tungsten with relaxed (110) surfaces calculated with 
the intersite matrix elements of the spin-orbit
Hamiltonian [Eq. (\ref{inter})] omitted from the calculation. 
Meaning of the colors is as in Fig. \ref{spin}. 
}
\label{nointersite} 
\end{figure*}

Fig. \ref{surf} shows the calculated energies $E$ of the subband 
Bloch states vs. in-plane wave vector for a 50 atomic layer 
thin film of tungsten with (110) surfaces for high symmetry cuts 
$\bar{\Gamma}{-}\bar{\text{H}}$
and $\bar{\Gamma}{-}\bar{\text{N}}$ through the 2D Brillouin zone 
that is displayed in Fig. \ref{BZN}(a).
The surface atomic layers of the film are relaxed towards the bulk 
by the experimentally
measured\cite{Meyerheim,Venus} 2.75\% of the bulk interlayer 
spacing. Surface states are
plotted in color. Those with probabilities $P$ of the electron 
being in the (110) surface atomic
layer are shown in red, orange and yellow for $P > 0.3$, 
$0.3 > P > 0.15$ and $0.15 > P > 0.075$,
respectively. The anisotropic upper and lower Rashba-Dirac 
cones with their common 
apex at $\bar{\Gamma}$
are clearly visible, superposed in color on the discrete subband 
structure of the thin film.
The energy splittings between the upper and lower cones are 
much larger in the 
$\bar{\Gamma}{-}\bar{\text{H}}$ direction than in the 
$\bar{\Gamma}{-}\bar{\text{N}}$ direction,
as expected from previous work on the surfaces of 
macroscopic crystals.\cite{Miyamoto2012a,
Miyamoto2012b,Rybkin2012,Biswas2013,Mirhosseini2013a,Giebels2013,
Mirhosseini2013b,
Braun2014,Miyamoto2015,Miyamoto2016} 

The strongly spin polarized surface states are shown in Fig. \ref{spin}
for a cut through the 2D Brillouin zone along the line 
$\bar{\text{H}}{-}\bar{\Gamma}{-}\bar{\text{H}}$ in the Brillouin zone. 
The states shown in color have
a strong presence on the (110) surface atomic
layer (as in Fig. \ref{surf}) and {\em also} are strongly spin polarized, i.e., 
more than 90\% spin $\uparrow$ or spin $\downarrow$ in the (110) surface
atomic layer. In Fig. \ref{spin}, the spin quantization axis is in the 
(110) plane and perpendicular to the wave vector
of the subband Bloch state. As expected for Rashba spin-split states\cite{BR1,BR2} and
consistent with experiments and previous theories for (110) surfaces of macroscopic 
tungsten crystals,\cite{Miyamoto2012a,
Miyamoto2012b,Rybkin2012,Biswas2013,Mirhosseini2013a,Giebels2013,
Mirhosseini2013b,
Braun2014,Miyamoto2015,Miyamoto2016} 
the spin polarization is opposite for the states of the upper and lower
Rashba-Dirac cones with the same in-plane $k$ vector and also 
opposite for states with the same energy and opposite $k$ vectors.  
The results for the opposite $(\bar{1}\bar{1}0)$ surface layer (not shown) 
are similar but the roles of
spin $\uparrow$ and $\downarrow$ are interchanged.

The results shown in Figures \ref{surf} and \ref{spin} agree very well 
with the findings of both experiments and {\em ab initio} 
calculations,\cite{Miyamoto2012a,
Miyamoto2012b,Rybkin2012,Biswas2013,Mirhosseini2013a,Giebels2013,
Mirhosseini2013b,Braun2014,Miyamoto2015,Wortelen2015,Miyamoto2016}
if one allows for the fact that the previous work has focused on 
the Rashba-Dirac cone states at the (110) surfaces 
of macroscopic tungsten
crystals that do not have the discrete subband structure of the thin films
considered here. 
However, the thin film character of the present system makes possible
new insights into the effects that hybridization with the bulk like states
has on the Rashba-Dirac cone surface states. In particular,
as is clearly
visible in Figures \ref{surf} and \ref{spin}, the strong hybridization with bulk-like  states
breaks up the upper
Rashba-Dirac cone in the $\bar{\Gamma}{-}\bar{\text{H}}$ direction into a series of
anticrossings. This results in the direction of group velocities 
$v = \frac{1}{\hbar}\frac{\partial E}{\partial k_{\bar{\Gamma}{-}\bar{\text{H}}}}$
of electrons in the states of the upper cone (except extremely close to the cone's apex) 
being {\em opposite} to that predicted
by considering only the over all dispersion
of the Rashba-Dirac cone without examining in detail the effects of hybridization
on individual electronic subband states. 

In view of the Rashba-like locking of the spin orientation 
to the direction of the 
wave vector of the
Rashba-Dirac cone surface states in Fig. \ref{spin}, this behavior of the group 
velocity is interesting from the
perspective of spintronics. This is because it means that in a thin film the directions of
travel of spin $\uparrow$ and $\downarrow$ surface state electrons can be the reverse
of the directions predicted from Rashba phenomenology and also from 
the $E(k)$ dispersion of surface states of {\em semi-infinite} crystals deduced 
from {\em ab initio} calculations (such as those in Ref. \onlinecite{Mirhosseini2013a})
of spectral densities of states in the Brillouin zone.
 
On close inspection, a similar reversal  relative to the over all dispersion
of the Rashba-Dirac cone is found
for the group velocities of electrons in states of the upper
Rashba-Dirac cone in the $\bar{\Gamma}{-}\bar{\text{N}}$ direction
in Fig. \ref{surf}.
However, hybridization has no such effect on the lower Rashba-Dirac cone 
in the $\bar{\Gamma}{-}\bar{\text{H}}$ direction; no anticrossings
are visible there in Figures \ref{surf} and \ref{spin}. Also while a series of anticrossings is
present in the lower Rashba-Dirac cone 
in the $\bar{\Gamma}{-}\bar{\text{N}}$ direction, the hybridization with the bulk
states there is extremely weak so that the anticrossings occupy only very narrow
ranges of $k$-space. The very different degrees of hybridization of the bulk-like
states with the upper and lower Rashba-Dirac cone surface states may be
expected to result in different intrinsic lifetimes of electrons in surface
states belonging to the upper and lower cones.

A similar calculation to that for Fig. \ref{surf} has been carried out
for a film with no surface
relaxation, i.e., the bulk structure of bcc tungsten was adopted without modification
throughout the 50 atomic layer film. The results are shown in Fig. \ref{BZN}(b)
for Bloch state energies and wave vectors near the location of the Rashba-Dirac 
cone apex of Fig. \ref{surf}. Only part of
the upper branch of the Rashba-Dirac cone in the 
$\bar{\Gamma}{-}\bar{\text{H}}$ direction is visible
in Fig. \ref{BZN}(b) where it is lower in energy by 
$\sim 0.1$eV than in Fig. \ref{surf}.  Evidently,
the surface relaxation is necessary for the Dirac cone apex, the lower Dirac cone
in the $\bar{\Gamma}{-}\bar{\text{H}}$ direction and both the upper and lower cones
in the $\bar{\Gamma}{-}\bar{\text{N}}$ direction 
to be present for the tight binding model developed here, as has 
also been the case for previous 
{\em ab initio}\cite{Mirhosseini2013a} calculations. 
In both the present tight
binding model and the previous {\em ab initio} calculations the surface 
relaxation opens an energy gap
between the apex of the Rashba-Dirac cones and the top of the closest 
energy range (marked in green in Fig. \ref{surf}) with an especially
high density of bulk states at the $\bar{\Gamma}$ point. 
In the present model the size of this gap is $\sim 0.06$eV for the relaxed structure,
a very similar value to those obtained from {\em ab initio} calculations, for example,
also $\sim 0.06$eV in Ref. \onlinecite{Mirhosseini2013a}. This gap is important
since if it were to close
or become negative (as in Fig. \ref{BZN}(b) for the unrelaxed surface structure)
the states of the Rashba-Dirac cones that fall in the energy range with the 
high density of bulk states would hybridize so strongly with the bulk states
as to be damped out, as in Fig. \ref{BZN}(b), and would no longer be observable 
as surface states.  

The degree of quantitative agreement between the above results for the
Rashba-Dirac surface states and 
both experiments and {\em ab initio} calculations for macroscopic
tungsten crystals\cite{Miyamoto2012a,
Miyamoto2012b,Rybkin2012,Biswas2013,Mirhosseini2013a,Giebels2013,
Mirhosseini2013b,Braun2014,Miyamoto2015,Wortelen2015,Miyamoto2016}
is quite remarkable since all of the parameters of the tight binding model
(described in Section \ref{M}) have been fitted {\em only} to the 
electronic band structure \cite{Christensen,Glantschnig,Papaconstantopoulos} 
of {\em bulk} bcc tungsten. The only information about the tungsten surface
that has been used as input in the present work is structural, 
namely, the experimentally measured
relaxation distance\cite{Meyerheim,Venus} of surface atomic layer towards the bulk.
No fitting of any kind to the properties of surface electronic states has been
carried out. This makes it possible to draw some previously
inaccessible conclusions about
the nature of the tungsten (110) Rashba-Dirac surface states and the mechanism 
responsible for them, based on what is and is not included in the Hamiltonian
of the present tight-binding model.

Importantly, 
the present tight binding model does {\em not} include electric fields that
are due to charge transfer between the tungsten and the vacuum near
the surface or between the bulk and surface layers, but despite this
the Rashba-Dirac cones that it yields agree very well with the results of
the {\em ab initio} calculations and with experiment. Therefore, 
although {\em ab initio} calculations indicate that such electric fields 
are present\cite{Braun2014} 
and they may, in principle, give rise to a Rashba effect,
the present work indicates that 
their effect on the properties of the W(110) surface states with
Dirac cone-like dispersion is at most minor. Similarly, the excellent
agreement between the predictions of the present model for
the very important upward shift in energy of the Rashba-Dirac surface states
due to surface relaxation with the predictions of the 
{\em ab initio} calculations\cite{Mirhosseini2013a} (as discussed above)
indicates that the same electric fields make an at most minor contribution
to this effect. The present model shows that this upward
shift is mainly due to the increase in the overlaps $O_{ij}$ between atomic orbitals
of the surface and bulk atomic layers of the tungsten that results from the 
surface relaxation and the associated
changes in the Hamiltonian matrix elements between these orbitals
through Eq. \ref{Kij}. 

In the tight-binding formulation\cite{Mireles} of the original Rashba model,\cite{BR1}
the only matrix elements of the spin-orbit Hamiltonian $H_\text{SO}$ that are considered
are those that connect orbitals with opposite spin on neighboring
sites. These give rise to the characteristic Rashba dispersion and spin polarization of
the electronic states. This intersite-spin-flip tight binding Rashba Hamiltonian  
was introduced\cite{Mireles} in work on spin precession in 
Datta-Das transistors\cite{Datta} involving gated nanowires formed from
2-dimensional electron gases (2DEG's) in semiconductors. It has more recently been
applied\cite{KM2} in studies of the quantum spin Hall effect in topological
insulators. The present tight-binding Hamiltonian includes both
intrasite [Eq.(\ref{intra})] and
intersite [Eq.(\ref{inter})] matrix elements of the spin-orbit Hamiltonian.
In view of the above it is of interest to investigate the roles that the 
intrasite and
intersite matrix elements of the spin-orbit Hamiltonian 
play in the formation of the Rashba-Dirac
cones of W(110) surface
states. This is addressed in Fig.\ref{nointersite} where the results are shown of the
same calculation as for Fig.\ref{spin} but with the
intersite matrix elements of the spin-orbit Hamiltonian set to zero. In Fig.\ref{nointersite},
as in Fig.\ref{spin}, surface states with more than 90\% spin $\uparrow$ or 
spin $\downarrow$ in the (110) surface
atomic layer are shown in color, using the same color scheme as in Fig.\ref{spin}.
The Rashba-Dirac cones of strongly spin polarized surface states are 
still present in Fig.\ref{nointersite} and the Rashba-Dirac cone dispersion is little changed
from that in Fig.\ref{spin}. However, for the states of the Rashba-Dirac cones
in Fig.\ref{nointersite} the probabilities of finding the electron
in the surface atomic layer are significantly lower than in the corresponding
states in Fig.\ref{spin} and the spin polarizations are typically also
not quite as strong. For some of the strongly spin polarized surface states not
belonging to the Rashba-Dirac cones, the probability for finding the 
electron in the surface layer is substantially larger (state $a$) or smaller (state $b$)
in Fig.\ref{nointersite} than in Fig.\ref{spin}. It follows that intrasite 
(and not intersite) matrix elements
of the tight-binding spin-orbit Hamiltonian are primarily responsible for 
the occurrence of Rashba-like
surface states at W(110) surfaces, although the intersite matrix elements
modulate some details quantitatively.  This finding also strongly suggests
that not only intersite matrix elements (as in Ref. \onlinecite{KM2}) 
but also intrasite matrix elements of the spin-orbit
Hamiltonian need to be included when considering Rashba-like effects
in theoretical work on topological
insulators, spin Hall phenomena and related topics. It also suggests that
the mechanism responsible for the Rashba-like dispersion and spin
polarization of W(110) surface states differs fundamentally from that
of the original Rashba effect
since that can be modelled\cite{Mireles} with
just intersite
matrix elements of the spin-orbit
Hamiltonian. This conclusion is consistent 
with that reached earlier\cite{Krasovskii} (regarding the relative
weakness of Rashba splittings due to charge distribution asymmetry)
 for surface states of other 
materials, although the reasoning used then\cite{Krasovskii} was entirely 
different 
and did not involve tight-binding models.
   
 \section{Conclusions} 
\label{Discussion}

In this article, a tight-binding model of bcc tungsten that includes spin-orbit
coupling has been introduced
and applied to study the properties of the Rashba-Dirac-like surface states
of (110) tungsten thin films. Although the model is fitted to only the bulk
electronic band structure of bcc tungsten, because it is based on extended H\"{u}ckel
theory it is flexible enough to account very well for the striking effect on the electronic 
surface states of the films of the relaxation of the
surface atomic layer of the tungsten towards the bulk. The results obtained are consistent with
the findings of previous {\em ab initio} calculations and experiments on the
surface states of bulk tungsten crystals. However, 
the present work has revealed that hybridization with bulk modes impacts
the states of the upper and lower Rashba-Dirac cones of W(110) surfaces 
in qualitatively different 
ways: While hybridization does not, for the most part, affect the direction of the group velocity 
of electrons in states belonging to the lower Rashba-Dirac cone, the group 
velocities of electrons in the states of the upper Rashba-Dirac cone are reversed by
hybridization relative to their direction predicted by Rashba phenomenology,
except in the immediate vicinity of the cone's apex. This implies
reversal of the directions of travel of  spin $\uparrow$ and $\downarrow$ 
surface electrons of the upper Rashba-Dirac 
 cone relative to those expected from the phenomenology.
In this article, the hybridization has also been shown to be much 
weaker for the lower Rashba-Dirac cone than for the 
upper Rashba-Dirac cone. The present work has also shown that electric fields that
are due to charge transfer between the tungsten and the vacuum near
the surface or between the bulk and surface layers do not 
significantly affect the tungsten (110)
Rashba-Dirac surface states. It has also shown the effect of 
the surface relaxation on the Rashba-Dirac
surface states to be due to the increased overlaps between atomic orbitals 
of the surface and neighboring layers resulting from the relaxation. 
It has also demonstrated that, unlike in tight-binding models
of the Rashba effect in semiconductor 2DEG's, intrasite (not intersite) matrix elements
of the spin-orbit Hamiltonian are primarily responsible for the formation of the tungsten (110)
Rashba-Dirac cones, and for
their dispersion and their spin polarization. This finding may have important implications
for theoretical modeling of topological insulators, spin Hall phenomena and related topics.

This research was supported by NSERC, CIFAR, Westgrid,
and Compute Canada.

{

\end{document}